\documentclass[useAMS,usenatbib]{mn2e}
\usepackage{longtable,color}
\usepackage{graphicx}
\usepackage{mathrsfs}
\usepackage{epsfig}
\usepackage{natbib}
\usepackage{color}
\usepackage{float}
\usepackage{amsmath}
\usepackage{times}
\usepackage{upgreek}
\usepackage[varg]{txfonts}
\usepackage{enumerate}
\usepackage{verbatim}
\usepackage{booktabs}
\usepackage{multirow}
\usepackage{amssymb}
\usepackage{placeins}

\title[Triplet alignment]{Galaxy Triplets Alignment in Large-scale Filaments}
\author[Rong et al.]{Yu Rong$^{1,2}$\thanks{E-mail: rongyua@ustc.edu.cn}, Jinzhi Shen$^{1,2}$, Zichen Hua$^{1,2}$\\
$^{1}$Department of Astronomy, University of Science and Technology of China, Hefei 230026, China\\
$^{2}$School of Astronomy and Space Sciences, University of Science and Technology of China, Hefei 230026, Anhui, China\\
}

\begin{document}
\maketitle

\begin{abstract}

Leveraging the datasets of galaxy triplets and large-scale filaments obtained from the Sloan Digital Sky Survey, we scrutinize the alignment of the three sides of the triangles formed by galaxy triplets and the normal vectors of the triplet planes within observed large-scale filaments. Our statistical investigation reveals that the longest and median sides of the galaxy triplets exhibit a robust alignment with the spines of their host large-scale filaments, while the shortest sides show no or only weak alignment with the filaments. Additionally, the normal vectors of triplets tend to be perpendicular to the filaments. {{The alignment signal diminishes rapidly with the increasing distance from the triplet to the filament spine, and is primarily significant for triplets located within distances shorter than $0.2$~Mpc$/h$, with a confidence level exceeding $20\sigma$.}} Moreover, in comparison to compact galaxy triplets, the alignment signal is more conspicuous among the loose triplets. This alignment analysis contributes to the formulation of a framework depicting the clustering and relaxation of galaxies within cosmological large-scale filament regimes, providing deeper insights into the intricate interactions between galaxies and their pivotal role in shaping galaxy groups.

\end{abstract}
\section{Introduction}

The hierarchical genesis of cosmic structures, spanning from small-scale galaxy pairs to large-scale galaxy clusters and groups, represents a fundamental facet in comprehending cosmic evolution. According to the hierarchical evolution theory of cosmic structures, galaxy clusters or groups form gradually through the amalgamation of individual galaxies from large-scale filaments. Initially, the formation of galaxy pairs and galaxy triplets (comprising three luminous galaxies) is anticipated, subsequently binding more galaxies through gravitational interactions, fostering the expansion of larger-scale and higher-mass galaxy clusters. While early studies regarded galaxy triplets as systems composed of a galaxy pair with an additional third member, {{recent investigations, utilizing triplet catalogues \citep[e.g.,][]{Karachentseva79,Trofimov95,O'Mill12,Argudo-Fernandez15,Costa-Duarte16}}}, have unveiled that the majority of galaxy triplets exhibit indications of prolonged dynamical evolution, with the member galaxies undergoing a dynamic co-evolution and presenting similar properties \citep{Chernin00,Duplancic13,Duplancic15,Hernandez-Toledo11,Vasquez-Bustos23}. Consequently, galaxy triplets should be viewed as an extension of small compact groups \citep{Duplancic15,Costa-Duarte16}.

{{Early observational studies have revealed that the preponderance of members in galaxy triplets are late-type galaxies, constituting approximately 70\%\--80\% of the total member count \citep{Karachentsev88,Karachentseva79,Hernandez-Toledo11}. Triplets comprised of three early-type members exhibit the smallest harmonic radii, signifying their general compactness compared to triplets with one or more late-type members \citep{Vasquez-Bustos23}. There is no predominant galaxy in triplets with respect to stellar population attributes such as global color, star formation rate, and stellar population \citep{Vasquez-Bustos23,Duplancic13,Feng15}. In comparison to galaxies in pairs and groups, triplet galaxies demonstrate intermediate colors and star formation activities, and also exhibit a preference for residing in moderately dense environments \citep{Duplancic20}.}}

{{The influence of galaxy triplets on galaxy properties and evolution has been brought to light. For instance, it has been observed that the star formation activity of galaxies is augmented in triplets, while the recurrent interactions in triplets can also lead to gas stripping, turbulence, and shocks that suppress star formation in the systems \citep{Duplancic20}. The interactions also have a strong effect on the nuclear activity and morphology of galaxies \citep{O'Mill12}. \cite{Duplancic15} suggests that the formation of elliptical galaxies through mergers in galaxy triplets may be favored by the configuration and dynamics of long-term evolved systems, and that the dynamics of different triplets may also vary \citep[e.g.,][]{Tawfeek19,Duplancic15,Costa-Duarte16}, thereby influencing the evolution of the member galaxies.}}

As the initial stages of group assembly, galaxy triplets also offer a unique opportunity to investigate the hierarchical nature of structure formation, enabling the observation of environmental influences shaping the initial galaxy groups and the exploration of mechanisms involved in early structure formation, such as tidal interactions, collisions, mergers, and relaxation processes within galaxy groups \citep[e.g.,][]{Argudo-Fernandez15,Vasquez-Bustos23}. Consequently, the study of galaxy triplets provides deeper insights into the intricate interactions between galaxies and large-scale environmental effects in structure formation.

Alignment analysis provides a potent methodology for scrutinizing and constraining scenarios of structure formation. At small scales, the spinning up of dark matter haloes along the overarching velocity field within a large-scale structure is induced by anisotropic collapse and tidal torques \citep{Codis12,Laigle15,Libeskind13,Ganeshaiah19}, resulting in the alignment of the spins of disk galaxies parallel to the elongation of the host filamentary structures \citep{Barsanti22,Kraljic21,Tempel13a,Tempel13b,Zhang15,Dubois14}. Elliptical galaxies, formed primarily through galaxy-galaxy mergers along the parent large-scale filament, exhibit a tendency to be elongated in the direction of the filaments after the conversion of the angular momenta of merging systems to the spins of elliptical galaxies \citep{Barsanti22,Kraljic21,Tempel13a,Rong16}. On larger scales, such as in galaxy clusters, the elongation of member galaxies aligns with the direction of the host large-scale filaments \citep{Lee15,Tempel15a}; the galaxies in the outer regions of dynamically young galaxy clusters maintain their original alignment within large-scale filaments \citep{Rong15b,Rong20}; the satellite galaxies in dynamically relaxed galaxy clusters tend to orient toward the central galaxy \citep{Rong15a,Rong19,Yagi16} due to tidal interaction. On an intermediate scale, the orientation of galaxy pairs has been observed to strongly correlate with their host filaments, particularly for loose pairs, suggesting that the formation of galaxy pairs occurs along the filamentary pattern \citep{Tempel15b}. However, the alignment of galaxy triplets with the large-scale filaments remains elusive. Consequently, a comprehensive understanding of the early stages of galaxy cluster formation within large-scale filamentary structures is still incomplete.

In this study, we aim to explore the alignment between galaxy triplets and filaments. Section 2 provides a concise introduction to the triplet sample and filament catalog utilized in our investigation. Section 3 presents our findings pertaining to the search for alignment signals. A comprehensive summary and discussion of our results will be presented in Section 4.

\section{Sample of galaxy triplets}

The galaxy triplet sample is derived from the galaxy group catalog established by \cite{Tempel12}, which is based on data from the Sloan Digital Sky Survey (SDSS) \citep{Aihara11}, and extracted using the friends-of-friends (FoF) grouping algorithm with a variable linking length. A galaxy group is classified as a triplet if it comprises only three (bright) member galaxies (apparent $r$-band magnitude limit is 17.77~mag). While a triplet may also include additional dwarf galaxy members, we disregard these faint objects due to their weak gravitational influence, which does not significantly impact the formation of galaxy groups. \cite{Tempel12} provides the properties of galaxy groups (triplets) and their member galaxies, including their positions, $ugriz$-band absolute magnitudes, and morphologies, etc.


For each galaxy triplet, we initially ascertain its associated large-scale filament from the filament catalog compiled by \cite{Tempel14a}. The filament identified as the host filament of the triplet is determined based on the minimum distance ($d_{\rm{tf}}$) from the geometric center of the triplet to the spine of a large-scale filament {{in the three-dimensional space.} In this study, we specifically investigate the alignment of galaxy triplets within a range of {{$d_{\rm{tf}}\leq 1.0$~Mpc$/h${\footnote{In section~\ref{sec3}, we will show that the alignment signal decreases very fast with increasing $d_{\rm{tf}}$, and the alignment signal has already disappeared when $d_{\rm{tf}}>0.2/h$~Mpc; therefore, the threshold of $d_{\rm{tf}}\leq 1$~Mpc$/h$ has been enough for studying the triplet alignment in large-scale filaments.}}. Ultimately, a total of 3,815 triplets and their respective host large-scale filaments are selected for analysis.}

Viewing each galaxy in a triplet as a vertex of the triangle, we can compute the vectors of the three sides (i.e., {$A_1$}, {$A_2$}, and {$A_3$}). To mitigate potential systematic effects introduced by the redshift-based distances, the three sides as well as the spine (denoted as {$S$}) of the host filament are then projected onto the plane of the celestial sphere to evaluate potential alignment signals in the projected plane.

For each galaxy triplet projected onto the celestial sphere plane, we measure the angles between the three projected sides ($A'_1$, $A'_2$, and $A'_3$, representing the projected sides for $A_1$, $A_2$, and $A_3$, repectively (as sketched in Fig.~\ref{sketch}) and the projected spine ($S'$) of the host filament, denoted as $\beta_1$, $\beta_2$, and $\beta_3$. The angles $\beta_1$, $\beta_2$, and $\beta_3$ are confined to the range of [0,90$^\circ$]. An alignment signal is identified if the distribution of $\beta_1$, $\beta_2$, or $\beta_3$ significantly deviates from a uniform distribution. We define a parameter $I(\alpha)=N_{0-45}/N_{45-90}$ to quantitatively indicate the strength of the alignment signal on the sky plane{\footnote {The error of ${\mathcal{I}}(\alpha)$ is estimated using the bootstrap method with 1,000 iterations. The standard deviation of ${\mathcal{I}}(\alpha)$ is then considered as the error.}}, where $N_{0-45}$ and $N_{45-90}$ represent the numbers of galaxies with angles $\alpha$ in the ranges of [0,45$^{\circ}$] and [45$^{\circ}$,90$^{\circ}$], repectively, with $\alpha$ representing $\beta_1$, $\beta_2$, or $\beta_3$. A uniform distribution corresponds to ${\mathcal{I}}(\alpha)\simeq 1$.

\begin{figure}
\centering
\includegraphics[width=\columnwidth]{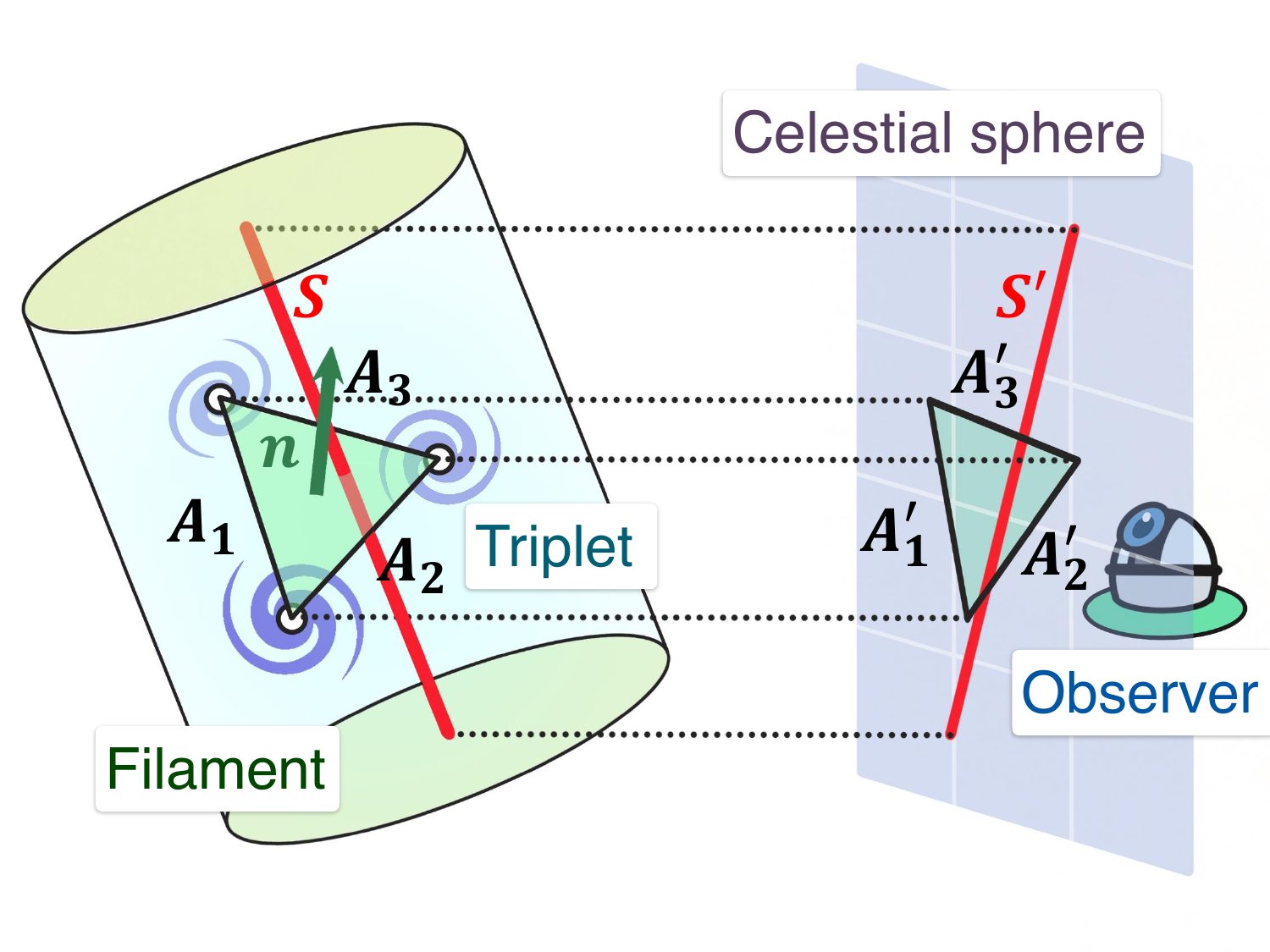}
\caption{A schematic representation of a galaxy triplet within a large-scale filament. The three sides of the triangular configuration are labeled as $A_1$, $A_2$, and $A_3$, while their corresponding projections onto the celestial sphere are denoted as $A'_1$, $A'_2$, and $A'_3$. The normal vector to the triplet plane is represented as $n$. The spine and its projection of the large-scale filament are indicated as $S$ and $S'$, respectively.}
\label{sketch}
\end{figure}


\section{Alignment of triplets in large-scale filaments}\label{sec3}

Firstly, we arrange the three sides of each triplet based on their respective lengths, strictly requiring $| A_1 |>| A_2 |>| A_3 |$, and then construct the distributions of $\beta_1$ (depicted in blue histogram), $\beta_2$ (depicted in green histogram), and $\beta_3$ (depicted in magenta histogram) for our triplet sample. As depicted in panel~a of Fig.~\ref{tri_alig}, we observe that the first and second longest sides of the triplets exhibit significant alignment signals at the $9\sigma$ confidence level, with ${\mathcal{I}}(\beta_1)=1.46\pm0.05$ and ${\mathcal{I}}(\beta_2)=1.39\pm0.04$. The longest sides $A_1$ display the most pronounced alignment with the filamentary structure, while the median sides $A_2$ show a moderate alignment signal. In contrast, the shortest sides $A_3$ demonstrate weak or negligible alignment signals on the sky plane (${\mathcal{I}}(\beta_3)=1.17\pm0.05$). 

We hypothesize that the strength of the alignment signal is associated with the mass of the galaxies corresponding to each side. Accordingly, we assess the stellar mass of the galaxy pair linked to each side and re-sort the three sides based on their ``mass weights'', where $A_1$ connects the most massive two galaxies in the triplets, and $A_3$ links the least massive two galaxies in the triplets. The stellar mass of a galaxy is estimated from the $i$-band absolute magnitude as well as the $g$ and $i$-band color $g-i$, utilizing the mass-to-light ratio of $\log_{10} (M_{\star}/L_i)=0.70(g-i)-0.68$ \citep{Taylor11}. We observe that the strength of the alignment signals remains largely unchanged with the mass weights, as indicated by the two-sample's Kolmogorov–Smirnov (K-S) test $p$-values (depicted in panel~b of Fig.~\ref{tri_alig}), as well as similar ${\mathcal{I}}(\beta)$ values, specifically ${\mathcal{I}}(\beta_1)=1.37\pm0.05$, ${\mathcal{I}}(\beta_2)=1.34\pm0.04$, and ${\mathcal{I}}(\beta_3)=1.30\pm0.04$. This suggests that the strength of the alignment signal is contingent solely upon the side lengths, rather than the mass weights. The mass weights may be unrelated to the side lengths in triplets. Consequently, it is evident that the sequence in which the three member galaxies entered a triplet may be independent of their masses.

We further explore the relationship between alignment and triplet properties, encompassing triplet compactness (quantified by the sum of the distances from the projected triplet's geometric center to the three member galaxies), mass (derived as the total stellar mass of the member galaxies), average color of the member galaxies, average specific star formation rates{\footnote{Specific star formation rates are obtained from the MPA-JHU DR7 release of spectrum measurements (https://wwwmpa.mpa-garching.mpg.de/SDSS/DR7/\#derived)}}, as well as morphology of member galaxies, etc. Upon dividing our triplet sample into two subsamples of equal size, we observe a notable discrepancy in alignment signals solely in the comparison of loose and compact galaxy triplets, as illustrated in panel~c of Fig.~\ref{tri_alig}. The alignment signals of the three sides (with the requirement $| A_1 |>| A_2 |>| A_3 |$) in the loose triplets are considerably stronger than those in the compact triplets. This outcome may imply that the primordial alignment weakens with the dynamical relaxation of galaxy triplets. Apart from compactness, the alignment dependence on other triplet properties is statistically insignificant.

We also directly quantify the angles, $\eta$, between the normal vectors of the triplet planes (i.e., {$n$}) and the spines of the host filaments in three-dimensional space. In theory, if the orientations of $n$ are independent of the spine orientations, $\cos\eta$ should be uniformly distributed in the range of [0,1]. However, as depicted in panel~a of Fig.~\ref{tri_nor}, we observe that $\cos\eta$ tends to approach 0 (i.e., ${\mathcal{I}}(\cos\eta)= N_{0-0.5}/N_{0.5-1}=1.36\pm0.04$, where $N_{0-0.5}$ and $N_{0.5-1}$ represent the counts of triplets with $\cos\eta$ falling within the ranges of [0,0.5] and [0.5,1], respectively), consistent with $\eta\sim 90^{\circ}$, indicating that, statistically, $n$ tends to be perpendicular to filamentary spines. Furthermore, the alignment of normal vectors predominantly occurs in the loose triplets (${\mathcal{I}}(\cos\eta)=1.61\pm0.07$); the orientations of normal vectors of compact triplets are either randomly distributed or weakly aligned with filaments (${\mathcal{I}}(\cos\eta)=1.16\pm0.05$), as demonstrated in panel~b of Fig.~\ref{tri_nor}.

\begin{figure}
\centering
\includegraphics[width=\columnwidth]{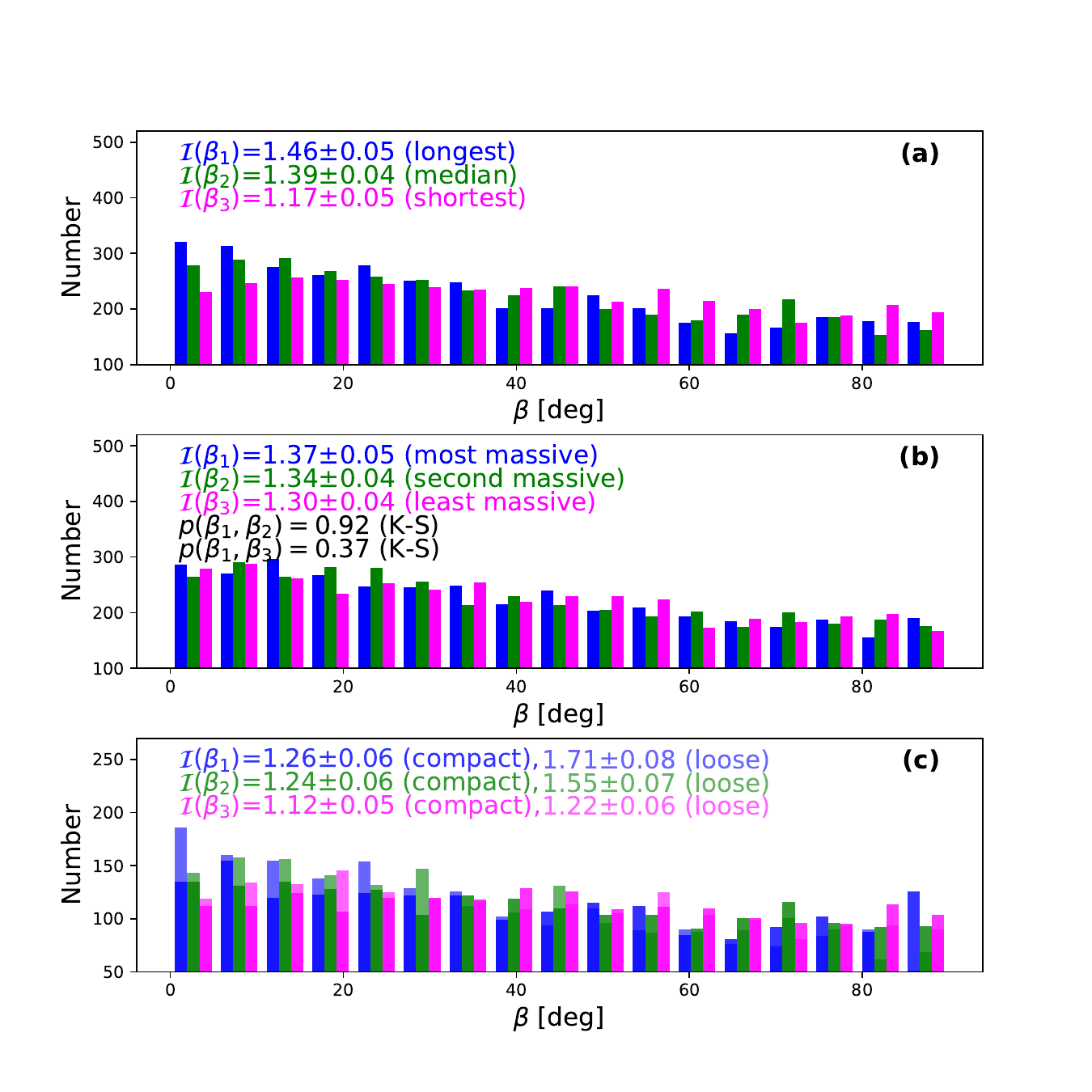}
\caption{ The distributions of $\beta_1$ (depicted in blue histogram), $\beta_2$ (depicted in green histogram), and $\beta_3$ (depicted in magenta histogram) {{for the triplet sample of $d_{\rm{tf}}\leq 1.0$~Mpc$/h$}}. Panel~a and b present the results of sorting the sides of triplets based on their lengths and mass weights, respectively. Panel c compares the distributions of $\beta_1$, $\beta_2$, and $\beta_3$ for compact (dark colors) and loose (light colors) triplets, with the sides of triplets sorted by lengths. The corresponding ${\mathcal{I}}(\beta)$ values are displayed. In panel b, $p(\beta_1,\beta_2)$ and $p(\beta_1,\beta_3)$ represent the $p$ values of the two-sample K-S test comparing the distributions of $\beta_1$ and $\beta_2$, as well as the distributions of $\beta_1$ and $\beta_3$, respectively.}
\label{tri_alig}
\end{figure}

\begin{figure}
\centering
\includegraphics[width=\columnwidth]{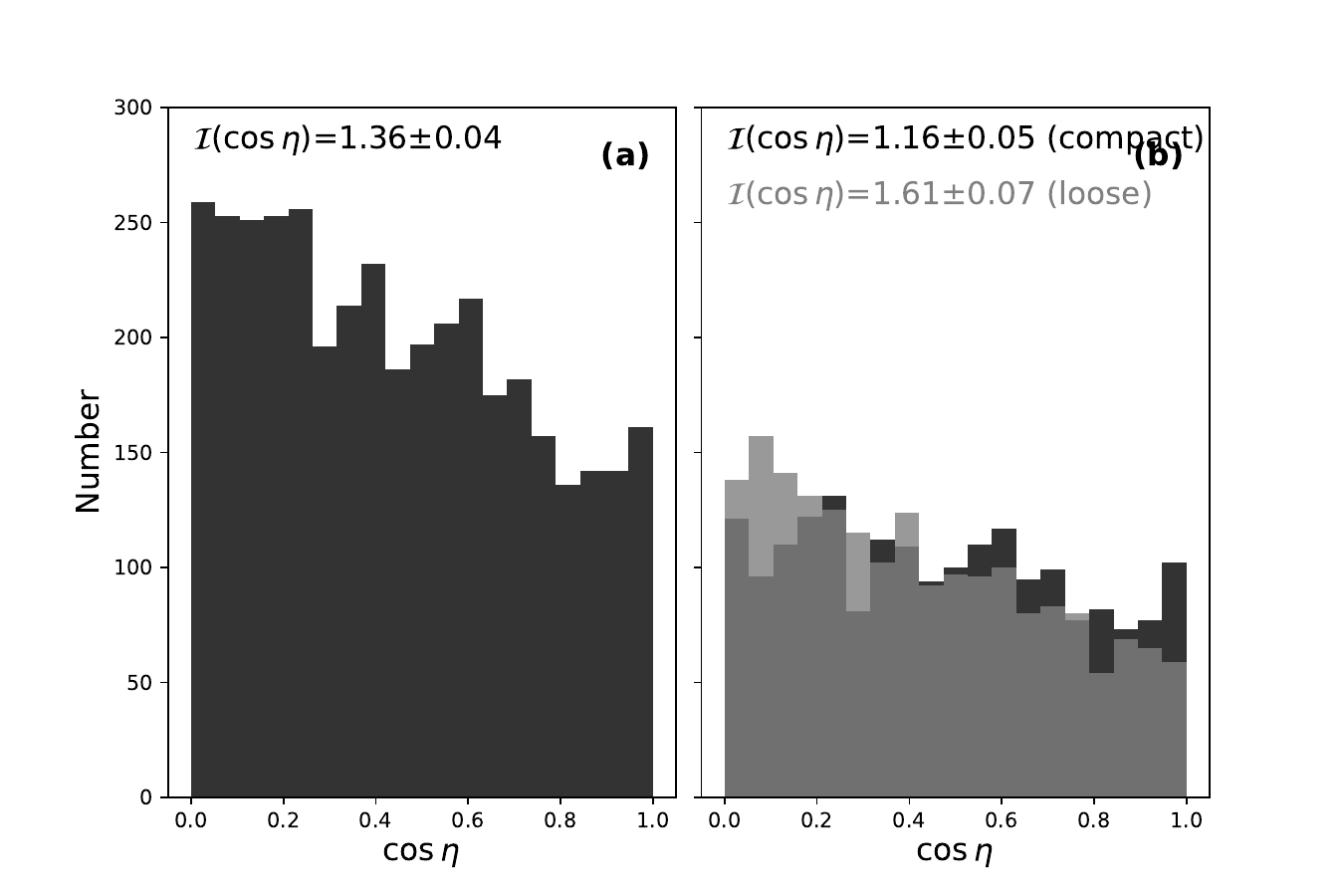}
\caption{The distributions of $\cos\eta$, representing the alignment between the normal vectors of triplet planes and their respective large-scale filaments { for the triplet sample with $d_{\rm{tf}}\leq 1.0$~Mpc$/h$}. Panel a illustrates the results for the entire triplet sample. Panel~b displays the distributions for the compact (black) and loose (grey) triplets, along with the corresponding ${\mathcal{I}}(\cos\eta)$ values for the compact and loose subsamples.}
\label{tri_nor}
\end{figure}

\begin{figure}
\centering
\includegraphics[width=\columnwidth]{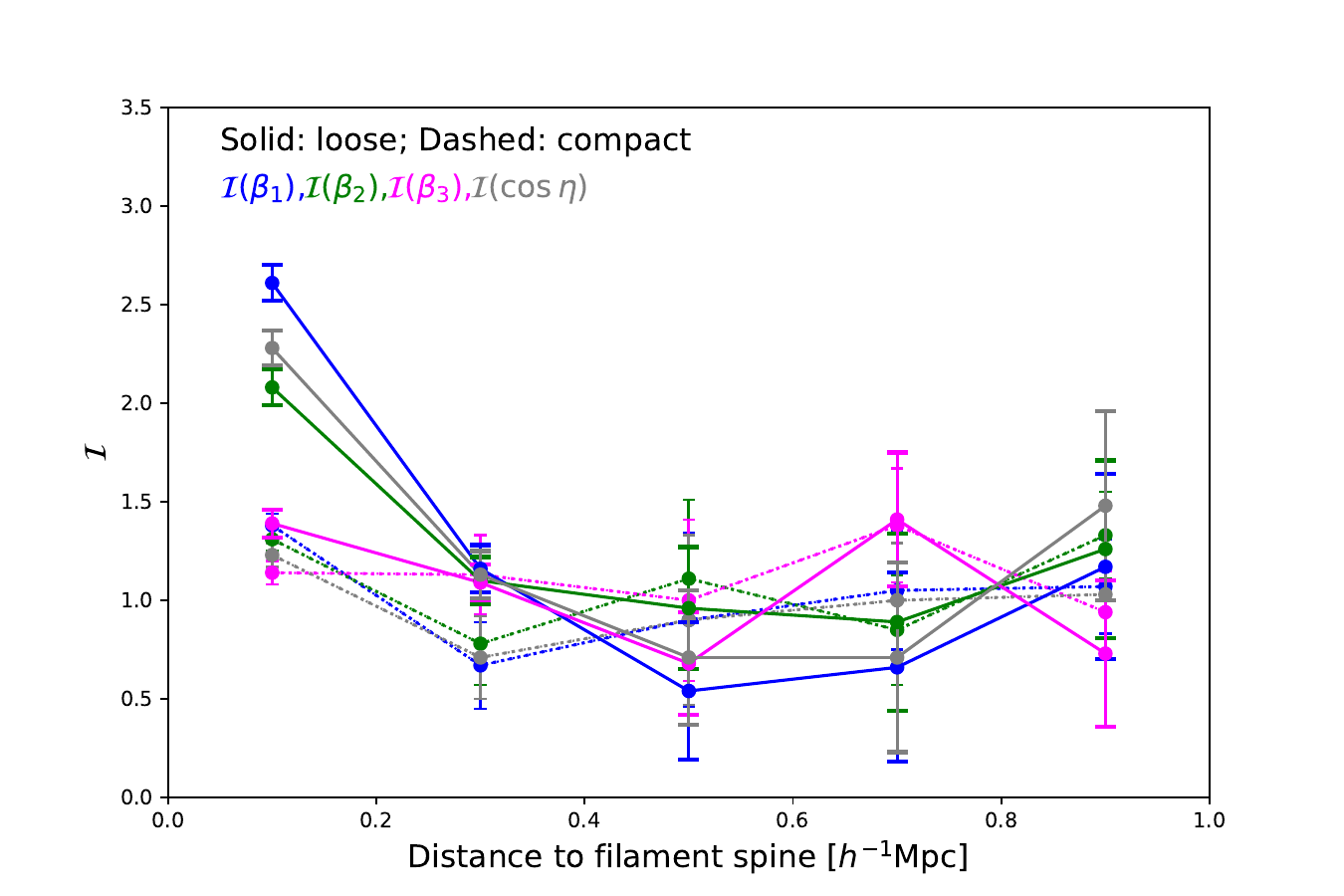}
\caption{The alignment signal strength of $\beta_1$ (blue), $\beta_2$ (green), $\beta_3$ (magenta), and $\cos\eta$ (grey), i.e., ${\mathcal{I}}(\beta_1)$, ${\mathcal{I}}(\beta_2)$, ${\mathcal{I}}(\beta_3)$, and ${\mathcal{I}}(\cos\eta)$, as a function of the distance to filament spine $d_{\rm{tf}}$. Here we impose the condition $| A_1 |>| A_2 |>| A_3 |$. The alignment signal strengths of the loose and compact triplets are visually differentiated by the solid and dotted components, respectively.}
\label{align_dis}
\end{figure}

{{Finally, given the sensitivity of the alignment signal to the proximity of large-scale filaments \citep[e.g.,][]{Wang20,Chen19}, we proceed to partition the triplet samples into distinct bins based on the distance within $d_{\rm{tf}}\leq 0.2$~Mpc$/h$, $0.2<d_{\rm{tf}}\leq 0.4$~Mpc$/h$, $0.4<d_{\rm{tf}}\leq 0.8$~Mpc$/h$, and $0.8<d_{\rm{tf}}\leq 1.0$~Mpc$/h$. Subsequently, we examine the strength of the alignment signal (as represented by ${\mathcal{I}}(\beta_1)$, ${\mathcal{I}}(\beta_2)$, ${\mathcal{I}}(\beta_3)$, and ${\mathcal{I}}(\cos\eta)$, when $| A_1 |>| A_2 |>| A_3 |$) as a function of $d_{\rm{tf}}$. As illustrated in Fig.~\ref{align_dis}, the alignment signal diminishes rapidly with increasing $d_{\rm{tf}}$; notably, only the loosely arranged triplets within $d_{\rm{tf}}\leq 0.2$~Mpc$/h$, i.e., those in close proximity to the filamentary spines, exhibit significant alignment. Conversely, those situated farther from the filament spines demonstrate negligible or very weak alignment with the large-scale filaments. The longest and median sides of the innermost triplets with $d_{\rm{tf}}\leq 0.2$~Mpc$/h$, as well as the normal vectors of the triplet planes, exhibit significant signals at confidence levels of $28\sigma$, $24\sigma$, and $25\sigma$ confidence levels, respectively\footnote{The confidence level is estimated as $({\mathcal{I}}-1)/e_{{\mathcal{I}}}$, where ${\mathcal{I}}$ denotes ${\mathcal{I}}(\beta_1)$, ${\mathcal{I}}(\beta_2)$, ${\mathcal{I}}(\beta_3)$, or ${\mathcal{I}}(\cos\eta)$, and $e_{{\mathcal{I}}}$ is the corresponding uncertainty estimated with the bootstrap method.}.}

\section{Summary and discussion}

Utilizing observational samples of galaxy triplets and large-scale filaments, we have investigated the alignment between triplets and their host filaments. Our findings reveal robust alignment signals between the triplet sides and filament spines, particularly for the first and second longest triplet sides. Consistent with the results of side alignments, the normal vectors to the triplet planes are more likely to be perpendicular to the filament spines. The alignment between triplet sides and large-scale filaments is a plausible outcome, as galaxies are more likely to be accreted to triplets along filaments.

{{The magnitude of the alignment signal between triplet sides and the spine of large-scale filaments diminishes rapidly as the distance from the triplet to the filamentary spine increases. The alignment signal exhibits statistical significance solely for triplets situated within a distance of $d_{\rm{tf}}<0.2$~Mpc$/h$, with the signal strength potentially exceeding a confidence level of $20\sigma$.}}

We have also observed that the alignments of sides and normal vectors are only pronounced in the loose triplets, in contrast to their compact counterparts. {{ In order to investigate whether the dependence of alignment signal strength on triplet compactness arises from the possibility that the loosely arranged triplets may be situated closer to the filamentary spines, we have calculated the median $d_{\rm{tf}}$ for both the loose and compact triplet samples. Our analysis reveals a smaller median for compact triplets (median $d_{\rm{tf}}=0.13_{-0.08}^{+0.92}$~Mpc$/h$), compared to the loose ones, with a median $d_{\rm{tf}}=0.21_{-0.11}^{+0.86}$~Mpc$/h$. This suggests that the disparity in alignment signal strength is indeed attributable to the difference in compactness.}} 

The disparity in alignment strength between compact and loose triplets may imply that the compact triplets are more dynamically relaxed and older{{, i.e., well-evolved systems, leading to the dilution of their primordial alignment with large-scale filaments as a consequence of interactions between member galaxies within the triplets. Conversely, the loose triplets may represent hierarchically forming structures, likely at a considerable distance from virialization.}}

{{This discovery aligns with prior investigations into the dynamical states of triplets by \cite{Trofimov95}, which indicated that galaxies within compact triplets are likely to have undergone multiple interactions and are potentially in a state of equilibrium within the system. Conversely, the loose (wide) triplets are presumed to be dynamically youthful and potentially distant from virialization.}}


Furthermore, we observe that the strength of the side alignment signal is not substantially influenced by other triplet properties, such as masses, colors, specific star formation rates, etc. {{This observation implies that triplets with varying dynamical ages may not manifest significant disparities in the characteristics of their constituent galaxies. Additionally, we have conducted a direct assessment of the alignment of triplet sides in three-dimensional space and have confirmed that our conclusions remain unaltered.}}

\section*{Acknowledgments}

Y.R. acknowledges supports from the NSFC grant 12273037, and the CAS Pioneer Hundred Talents Program, as well as the USTC Research Funds of the Double First-Class Initiative.

\section*{Data Availability}

Data are available if requested.

\bibliographystyle{mn2e}


\end{document}